\documentclass[conference]{IEEEtran}
\IEEEoverridecommandlockouts
\usepackage{amsmath,amssymb,amsfonts}
\usepackage{algorithmic}
\usepackage{graphicx}
\usepackage{textcomp}
\usepackage{xcolor}       
\usepackage{algorithm}
\usepackage{geometry}

\def\BibTeX{{\rm B\kern-.05em{\sc i\kern-.025em b}\kern- .08em
    T\kern- .1667em\lower.7ex\hbox{E}\kern- .125emX}}

\begin{document}

\title{Study of Weighted Residual Layered Belief Propagation for Decoding of LDPC Codes \\
{\footnotesize 
}
\thanks{This work was supported by CAPES, CNPq and FAPERJ.}
}

\author{Hassan Touati and Rodrigo C. de Lamare \\
CETUC, DEE, PUC-Rio, Rio de Janeiro, Brazil  \\
University of York, UK\\
Email: touati.hassan@hotmail.com and rodrigo.delamare@york.ac.uk
}

\maketitle

\begin{abstract}
In this work, we investigate the decoding of Low-Density Parity-Check (LDPC) codes using informed dynamic scheduling algorithms that require a reduced number of iter- ations. In particular, we devise the weighted residual layered belief propagation (WR-LBP) decoding algorithm, which exploits the residual within a structured layer framework to speed the number of required decoding iterations. The proposed WR-LBP algorithm is assessed against important LDPC decoding algorithms, in terms of the number of iterations required for convergence and the bit error rates. 
\end{abstract}


\section{Introduction}
The concept of low-density parity-check (LDPC) codes was introduced by Robert Gallagher in his doctoral dissertation thesis in 1960 \cite{b1}, however, at that time, LDPC codes did not gain widespread attention. LDPC codes can increase channel capacity by using a belief propagation decoding algorithm, which exchanges decoding messages iteratively between variable and check nodes. The potential of LDPC codes for error correction appeared after a long time in the 1990s by Mackay and Neal \cite{b2}. LDPC codes, characterized by sparse parity check matrices due to their linear block code nature, have only a small fraction of non-zero elements. This property enhances their suitability for transmission and storage channels. Additionally, LDPC codes and their improved versions \cite{peg,bfpeg,dopeg,memd,armo,baplnc,rsbd} are widely employed in modern communication systems, particularly in wireless communication standards such as Wi-Fi (IEEE 802.11n, 802.11ac), 4G LTE, and 5G NR.

In LDPC decoding, scheduling relates to the sequence and timing of message updates between variable nodes and check nodes inside the factor graph representation. The choice of scheduling technique affects the convergent speed, error correction performance, and overall complexity of the decoding algorithm. Informed dynamic scheduling (IDS) techniques \cite{b3,ids,kaids} have an important role in the context of belief propagation (BP) decoding \cite{b4} for LDPC codes. Noted for their capacity-approaching performance, they depend on iterative decoding processes, and the scheduling of these iterations strongly impacts the efficiency and effectiveness of the decoding algorithm used with a few iterations to have better performance than flooding BP. In this context, the Silent-Variable Node-Free RBP (SVNF)  \cite{b5} algorithm tackles the issue of silent variable nodes by sequentially selecting and updating each C2V message with the highest residual generated by the check node connected to each variable node. Residual Belief Propagation (RBP) and Residual-Decaying-Based Residual Belief Propagation (RD RBP 0.9)  \cite{b6,b7},\cite{RBP2} algorithms reduce the decoding algorithm's greediness by gradually decaying the updated C2V residual, reducing the likelihood of selecting an edge in dynamic decoding algorithms. Whereas, in the reliability-based residual belief propagation (Rel RBP) \cite{b8} and \cite{b9}, the update of the node with the greatest advantage occurs by assessing the reliability of the Log-Likelihood Ratios (LLRs) exchanged during message passing. Additionally, a distinct measure is introduced for each iteration. Uniformly Reweighted Belief Propagation (URW BP) \cite{wymeersch,b10} aims to improve convergence speed and error correction capabilities through systematic reweighting of messages. The Layered Belief Propagation (LBP) \cite{b11} algorithm is particularly effective for decoding large graphical models in error-correcting LDPC codes.

In this paper, we introduce an enhanced IDS strategy, termed the Weighted Residual Layered Belief Propagation (WR-LBP) algorithm. The main idea behind the proposed WR-LBP algorithm is to employ a weighted residual strategy in the LBP algorithm  This innovative algorithm showcases notable advancements, exhibiting diminished complexity and enhanced performance in comparison to conventional BP techniques founded on IDS methodologies. The distinctive weighting mechanism embedded within WR-LBP significantly enhances its efficiency, leading to a reduction in computational cost while concurrently achieving superior performance outcomes when benchmarked against established algorithms in the domain.

The succeeding portions of this work are structured as follows. In Section II, we present a succinct overview of the system model and articulate the problem statement. Section III offers our suggested method targeted at minimizing the complexity and latency inherent in various variants of LDPC codes through the deployment of our WR-LBP decoding algorithm. The subsequent discussion in this section comprises simulation findings across multiple IDS settings. In Section IV, we assess the proposed WR-LBP and competing algorithms using simulations and in Section V we draw concluding remarks about this work.

\section{System model and problem statement } 

In a digital communication system, several key elements collaborate to enable effective information transfer. Commencing with the raw data, referred to as information, it undergoes encoding at the source through a source encoder.

\begin{figure}[h]
\centering
\includegraphics[width=8.5cm]{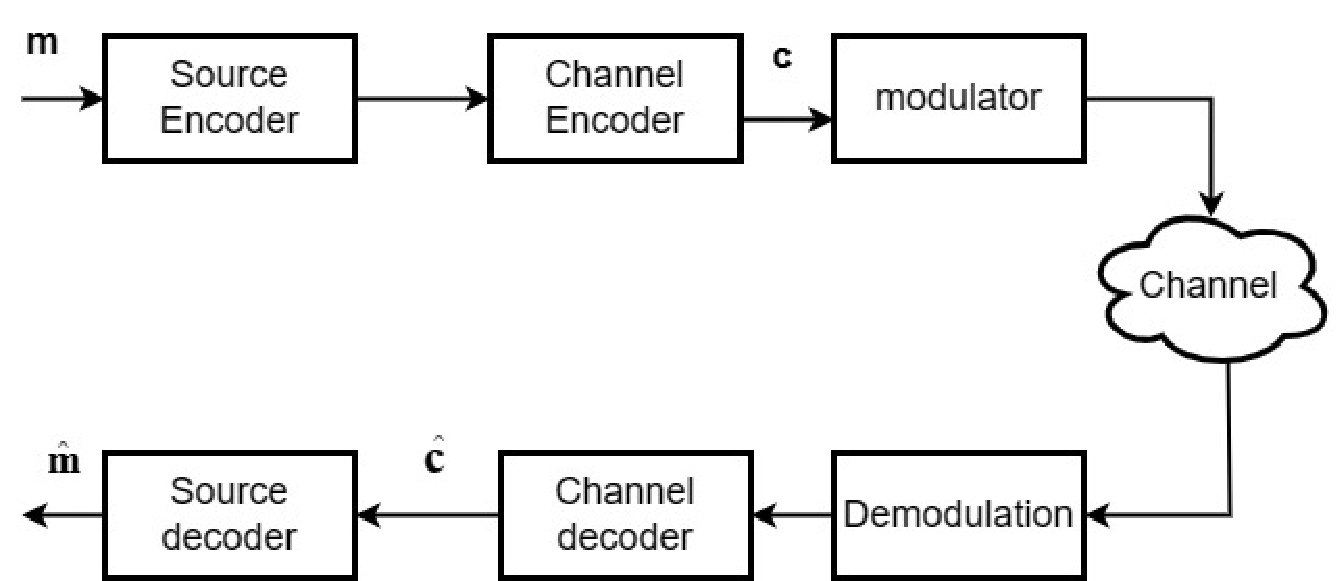}
\caption{Basic elements of a digital communication system.}
\renewcommand{\figurename}{Figure}
\renewcommand{\footnotesize}{\fontsize{8}{10}\selectfont}
\end{figure}

This encoded data, along with additional redundancy for error correction, is further processed by the channel encoder. The modulator then translates the digital signal into a suitable modulation form such as binary-phase shift-keying (BPSK) for propagation through the channel, as depicted in Fig. 1. As the signal is transmitted over the channel, it encounters noise and distortion. At the receiving end, the demodulator extracts the transmitted signal, and the channel decoder corrects errors introduced during transmission.  The reconstructed output, created using the source encoder, completes the repetitive cycle of digital communication. Each element, represented in Fig. 1, plays a key role in guaranteeing the correctness and reliability of information transmission in the communication system.

Generally, the mathematical description of the communication system in \cite{b12} Fig. 1 is given
\begin{align*} 
\mathbf{y}=\mathbf{c}+\mathbf{n}, \tag{1}
\end{align*} 
where $\mathbf y$ is the $n$-dimensional received vector, $\mathbf c$ is the $n$-dimensional codeword and $\mathbf n$ is $n$-dimensional noise vector that is assumed to be Gaussian with zero mean and variance $\sigma^2$.  

The code $\mathbf{c}$ must be generate by: 
\begin{align*} 
\textbf{c}=\textbf{m}\times{\textbf{G}}, \tag{2}
\end{align*}
to be valid if it satisfies the syndrome calculation $s= \textbf{cH}^T $.
where $\mathbf m$ is the $k$-dimensional message vector with $R=k/n <1$ being the code rate, $\mathbf G$ is the $k \times n$ generator matrix and $\times$ denotes the modulo-2 multiplication.

Arranging the parity check matrix of an LDPC code in systematic form, we have
 \begin{align*} 
 \mathbf{H}_{sys}=[\mathbf{I}  
  \ \vert{ \ \mathbf{P}_{(n-k) \times k }}] \tag{3} \end{align*}
where $\mathbf{P}$ is the $(n-k)\times k$ parity matrix and \textbf{I} is an identity matrix. Rearranging the systematic parity check matrix, we get the generator matrix $\mathbf{G}$: 
\begin{align*} 
\mathbf{G}=[\mathbf{P}_{(n-k) \times k}^T \ \vert \   \mathbf{I}] \tag{4}
\end{align*}
The multiplication are GF(2), We can verify our results as :
\begin{align*} \textbf{G} \times \textbf{H}^T= \textbf{0} \tag{5}\end{align*}
LDPC codes are often represented by the $n-k \times n$ parity-check matrix $\mathbf{H}$, which is represented by
\[
\mathbf{H} = \begin{bmatrix}
    h_{11} & h_{12} & \cdots & h_{1n} \\
    h_{21} & h_{22} & \cdots & h_{2n} \\
    \vdots & \vdots & \ddots & \vdots \\
    h_{(n-k)1} & h_{(n-k)2} & \cdots & h_{(n-k)n},
\tag{6} \end{bmatrix}
\] 
where $h_{ij}$ are the elements of $\mathbf{H}$ from the field over which the code is defined, 'commonly the binary field GF(2) for binary codes'. $\textbf{H}$ is carefully designed based on the properties of the specific linear block code. 

\begin{figure}[h]
\centering
\includegraphics[width=6.5cm]{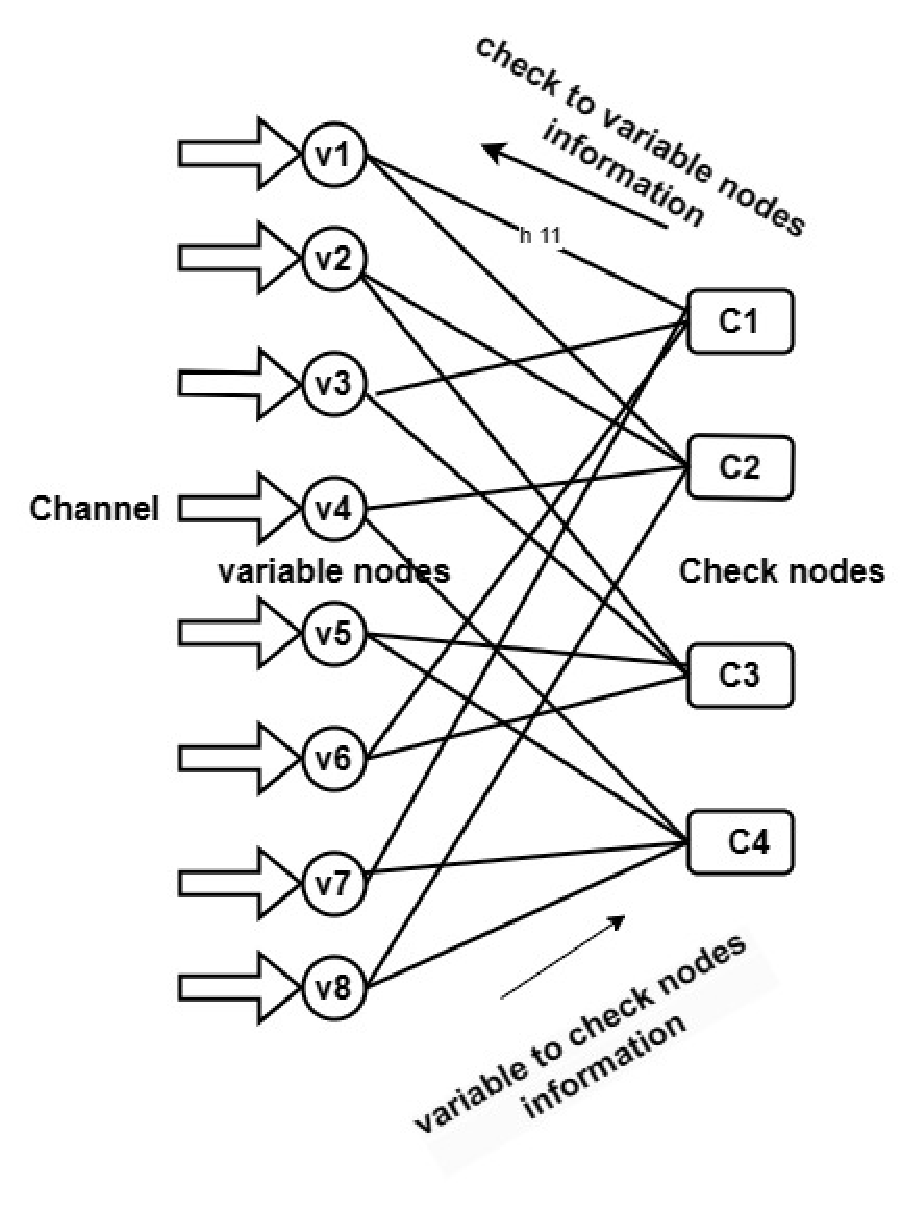}
\caption{Tanner Graph, exchanges intrinsic information between check nodes to variable nodes}
\end{figure}

The parity-check matrix $\mathbf{H}$ under a bipartite graph shown in Fig.2. can be represented in the following way:
The nodes are of two types: variable nodes represent the symbols of the codeword $\mathbf{c}$, which are circles, and check nodes correspond to the parity equation, which are squares. A check node $c_i $ is connected to a variable node $c_j $ when the element of $h_{ij} $ is
a one. The quantity $\mu_{c_{i}\rightarrow v_{j}}$  represents a message from the check node to the variable node. We also assume $\mu_{v_{j}\rightarrow c_{i}}$ is the message sent from a variable node to the check node. The indices $v_{j}$ and $c_{i}$ might refer to specific nodes in a graphic model.

A great deal of research currently focuses on improving LDPC decoding algorithms to enhance communication system performance and reduce latency. Specifically, it aims to create an intelligent scheduling strategy for LDPC decoding, limiting repeated updates for precise error correction. The goal is to reduce processing overhead and delay by strategically prioritizing the decoding process. This work aims to address LDPC decoding subtleties and contribute to advancing error correction techniques in communication systems.

\section{Proposed WR-LBP Decoding Algorithm}



In this section, we review the standard BP algorithm that is widely used for LDPC decoding and then present the proposed WR-LBP algorithm.

\subsection{Flooding Belief propagation}

The BP algorithm, also known as the sum-product algorithm (SPA) \cite{b13}, is a method used to compute the messages passed between variable nodes and check nodes, or vice versa, in the Tanner graph (see Fig. 3.). We assume that a codeword $\mathbf{c}_i=(c_1,c_2,....,c_N),$ is transmitted using BPSK modulation. This sequence is sent through a channel Additive White Gaussian Noise (AWGN) and the codeword estimated is $\hat{\mathbf{c}}_i=(\hat{c}_1,\hat{c}_2,....,\hat{c_N}) $.
\\

The BP algorithm is decomposed into four steps:
The first step  involves initializing the variable nodes by  the value of the log likelihood ratio (LLR) ${{L}}_j$ of the received sequence $\textbf{y}_j $ at each iteration given by
\begin{align*} 
{{L}}_j = \log\left(\frac{P(c_i=1|\textbf{y}_j)}{P(c_i=0|\textbf{y}_j)}\right),\tag{7}
\end{align*}
where this step can accelerate decoding by verifying the syndrome $ \mathbf{s}=\mathbf{c}\mathbf{H}^T $ where $\mathbf{H}$ is a parity-check matrix and $\mathbf{c}$ is the codeword. If $\mathbf{s}=0$ then the codeword is found and the process ends.

If not, the process continues to the second step, "Row Processing” where the computed message $\mu_{c_{i}\rightarrow v_{j}}^{(k)} $  at the check nodes is transmitted to the variables nodes:
 \begin{align*}
\resizebox{0.5\textwidth}{!}{%
$\begin{aligned}
\mu_{c_{i}\rightarrow v_{j}}^{(k)} &= \log\left({\frac{1+\tanh^{-1}\left(\prod_{\mathrm{j'}\in N(\mathrm{c}_{\mathrm{i}})\backslash \mathrm{j}}\tanh\left(\frac{\mu_{\mathrm{v}_{\mathrm{j}'} \rightarrow \mathrm{c}_{\mathrm{i}}}^{(\mathrm{k})_{}}}{2}\right)\right)}{1-\tanh^{-1}\left(\prod_{\mathrm{j'}\in N(\mathrm{c}_{\mathrm{i}})\backslash \mathrm{j}}\tanh\left(\frac{\mu_{\mathrm{v}_{\mathrm{j}'} \rightarrow \mathrm{c}_{\mathrm{i}}}^{(\mathrm{k})_{}}}{2}\right)\right)}}\right),
\end{aligned}$%
} \tag{8}
\end{align*}

From (8) to (9), we use the mathematical rule $\log(\frac{1+x}{1-x})=2\tanh^{-1}(x)$, and we obtain
\begin{align*} \mu_{c_{i}\rightarrow v_{j}}^{(k)}=2\tanh^{-1}\left(\prod_{\mathrm{j'}\in N(\mathrm{c}_{\mathrm{i}})\backslash \mathrm{j}}\tanh\left(\frac{\mu_{\mathrm{v}_{\mathrm{j}'} \rightarrow \mathrm{c}_{\mathrm{i}}}^{(\mathrm{k})_{}}}{2}\right)\right),\tag{9}\\ \end{align*}
The Third step is to compute the message from the variable nodes to the check nodes $\mu_{v_{j}\rightarrow c_{i}}^{(k+1)} $ by 
\begin{align*} \mu_{v_{j}\rightarrow c_{i}}^{(k+1)}= {L}_{j}+\sum_{i'\in N(v_{j})\backslash i}\mu_{c_{i'}\rightarrow v_{j}}^{(k)}\ , \tag{10} \end{align*}

At the end of each iteration, the variable node computes the total LLR of $M_j$ given by

\begin{align*} M_j^{(k+1)}={L}_{j}+\sum_{i\in N(v_{j})}\mu_{c_{i}\rightarrow v_{j}}^{(k)}\ , \tag{11} \end{align*}

In step four, the reliability value is calculated by equation (11) then the transmitted bits $ (\hat c, j=1, 2,…, N)$ are estimated from the hard decision according to  
\begin{align*}
                \hat c =  
\begin{cases}
  1 & \text{if $M_j$ \textless 0}, \\
  0 & \text{otherwise}
\end{cases}
, \tag{12} \end{align*}

If the syndrome is zero then the parity equation is satisfied, and the decoding stops and gives the codeword $\hat {\mathbf c}$ as a result. If not, the algorithm will be repeated, returning to step 2 until a valid codeword is reached or until the maximum number of iterations is reached. This terminates the decoding process.

\subsection{Weighted Residual Layered Belief Propagation algorithm}

  The RBP algorithm, as explained in reference \cite{b3}, \cite{b14}, \cite{b15}, operates by processing messages passed from check to variable nodes. This strategy, referred to as residual passing, involves examining the difference between newly updated messages and previous versions, In the binary RBP decoder, the selection of the next message to be updated is determined by the message residuals. In the RBP algorithm,  $ r(\mu_{c_{i}\rightarrow v_{j}}^{(k)})$ is the residual used and the message $\mu_{c_{i}\rightarrow v_{j}}^{(k)}$ is the newly update for check to variable nodes, whereas $\mu_{c_{i}\rightarrow v_{j}}^{(k-1)}$ is the previous update from the check to the variable nodes. The residual is written in terms of the messages according to
\begin{equation*} 
r(\mu_{c_{i}\rightarrow v_{j}}^{(k)})=\left\vert \mu_{c_{i}\rightarrow v_{j}}^{(k)}-\mu_{c_{i}\rightarrow v_{j}}^{(k-1)}\right\vert. \tag{13} \end{equation*}
The main notion of the RBP algorithm (13) is to prioritize the propagation of messages from check nodes to variable nodes, focusing on those where the residual has the most important impact on decoding. 
\begin{algorithm}[H]
    \label{algor1}
    \begin{algorithmic}[1]
        \caption{ WR-LBP Algorithm}\label{alg:cap}
        \STATE \textbf{Initialization}:\\Message from check to variable nodes:\\$ \mu_{c_{i}\rightarrow v_{j}}^{(0)}=0$ \\
        Message from variable to check  nodes:\\$\mu_{v_{j}\rightarrow c_{i}}^{(0)} = \mathbf{C_v} = -\frac{4 \cdot \mathbf{rx}}{\text{N0}}$ \\

        \STATE Initialize Variables and Matrices:$\textbf{R}_{cv}= 0_{M \times N }$.\\
        \textbf{Layer Assignment}:
        \STATE Before the horizontal step, the code calculates a layer assignment for each row of the LDPC matrix $\textbf{H}$.\\
        \STATE determines the highest level of recoverability $L$ among the variable nodes in each row.\\
        \STATE  Determine layer Assignment $M_L$.\\
            \textbf{Horizontal Step:} \\
           \STATE  \textbf{For} each  $iteration $ row in parity matrice $\textbf{H}$:  \\
                  \STATE    Updated messages from check nodes to Variable nodes by Equation $(8)$ or $(9)$.  \\
                
                   Multiply the result of updated messages passing in check nodes, by Layer Assignment $ M_L$.  \\
                    \STATE   Update residual matrix.
                    $ r(\mu_{c_{i}\rightarrow v_{j}}^{(k)})$  \\  
 \STATE \textbf{End For}\\
                      \textbf{Iteration:}\\ 
         \WHILE{the stopping rule is not satisfied}
           \STATE find maximum value in $\textbf{R}_{cv}$\\
        \ENDWHILE
    \STATE   For the selected check node, cb, identify and calculate the next-largest residual $  r_{\text WR-LBP}(\mu^{(k)}_{ v_{c} \rightarrow  c_{d}})$ and record the message $M^{(k+1)}_{c_b \rightarrow v_p}$ .\\
              \STATE  Update each 
        $\mu^{(k+1)}_{ v_{c} \rightarrow  c_{d}}$        ,where $d\in N(\mathrm{c})\backslash \mathrm{b}$                
        and messages passing from variable node to check node according to $(10)$ and $(11)$ respectively, and set the appropriate position in each $i_d $ to one. \\
        \STATE \textbf{for} each $e \in N(\mathrm{d})\backslash \mathrm{c}$ \textbf{do} \\
Perform the indicator vector check as above.
Identify and calculate the check node residual
 $ r_{\text WR-LBP}(\mu^{(k+1)}_{e})$ and record the message $\mu_{c_{e}\rightarrow v_{q}}^{(k+1)}$
, storing the value $q$ also.

\textbf{end for}

 \textbf{Decoding Condition:}\\
 \STATE  If the number of messages passed  is divisible by the total number of edges , indicating the completion of an iteration:\\
 \STATE  The decoded vector is calculated as the result of a comparison between the variable node messages $m_v$ and zero, converted to a double vector.\\
  
\STATE  \textbf{End Function;}
\STATE The algorithm incorporates auxiliary functions, including level Of Recoverability and layer Weight computed by: $ w(i)=\frac{1}{2^{(l-level+1)}}$ , to manage the decoding process effectively.\\

    \end{algorithmic}
\end{algorithm}
 This prioritizing helps quicker convergence times, particularly when these messages from check nodes to variable nodes are relayed first. \\
The proposed WR-LBP algorithm employs the residual concept of the RBP algorithm in the layered structure to the message-passing procedure, nodes are structured into layers, thereby facilitating a sequential exchange of messages between adjacent layers. This type of structure helped to improve the convergence speed and overall efficiency of the decoding process. Firstly, before the horizontal step, we determine a layer assignment for each row of the LDPC matrix $\mathbf H$ by  
\begin{align*} 
{M}_L = { \max}(L(\mu_{c_{i}\rightarrow v_{j}}^{(k)}),l), \tag{14}
\end{align*}
where $L$  is the high level of recoverability among variable nodes in each row.
The layer corresponding to the residual to be exchanged is given by
\begin{align*}
\resizebox{0.5\textwidth}{!}{%
$\begin{aligned}
L(\mu_{c_{i}\rightarrow v_{j}}^{(k)}) &= \min\left(\left[\frac{(l+1)}{2} \times ((1+\text{sign}(\mu_{c_{i}\rightarrow v_{j}}^{(k)})),l)\right]\right), \\
\end{aligned}$%
} \tag{15}
\end{align*}
where $sign$ is the sign function that returns -1 for negative values, 0 for zero, and 1 for positive values, the result truncated to be within the range [0 1] and  $l$ is a constant determined through empirical simulation or optimization we use the value 0.9.

The messages propagate from check to variable nodes, which was computed in the proposed WR-LBP algorithm by multiplying the residual belief propagation (RBP) in equation (13) by the layer assignment $M_L$, which results in 
\begin{equation*} 
r_{\text WR-LBP}(\mu_{c_{i}\rightarrow v_{j}}^{(k)})= \alpha_{({M}_L)} \times \left\vert \mu_{c_{i}\rightarrow v_{j}}^{(k)}-\mu_{c_{i}\rightarrow v_{j}}^{(k-1)}\right\vert , \tag{16} \end{equation*}

where $\alpha_{({M}_L)}$ is a constant based on $ {M}_L$ and determined by layer Weight $\frac{1}{2^{(l-M_L+1)}}$ .
 
\subsection{Computational complexity analysis}

One of the primary drivers behind the novel decoding algorithm WR-LBP introduced in this study was to address the significant computational burden imposed by existing IDS schemes when dynamically selecting each message for updating.

We assume that the total number of edges in the Tanner graph is given by $E = \hat{d_v} \times N = \hat{d_c} \times M$, where $N$ represents the variable nodes and $M$ represents the check nodes, with $\hat{d_c}$ being the average number of check nodes and $\hat{d_v}$ being the average number of variable nodes. All algorithms under consideration maintain the same number of updates from check nodes to variable nodes, denoted as $E$, and also feature an identical number of updates from variable nodes to check nodes, specifically $E(\hat{d_v}-1)$, except for BP and VFAP algorithms, which have $E$ updates in this regard. The pre-computation requirement for BP, VFAP, and LBP algorithms is $0$, whereas for other algorithms, it amounts to $E(\hat{d_v}-1)(\hat{d_c}-1)$. Similarly, the comparison process for BP, VFAP, and LBP algorithms incurs no pre-computation cost ($0$), while for the remaining algorithms, it results in a cost of $E(E-1)$, except for SVNF, where it is $E(\hat{d_v}(\hat{d_c}-1)-1)$.

The complexity of the algorithms outlined in this paper is detailed in Table I, indicating the frequency of utilization for both the variable and check node update equations. Notably, the figures in the table denote the usage of each update equation per standard iteration for every scheduling scheme. In the case of the modified iteration, each scheme uniformly employs the check node update equation a consistent number of times. The proposed approach can be also applied to iterative detection and decoding schemes in wireless communications \cite{jidf,spa,mfsic,dfcc,mbdf,did,bfidd,1bitidd,dynovs,msgamp1,msgamp2}.

\begin{table*}[htbp]
\caption{Complexity of decoding Algorithms}
\begin{center}
\setlength{\tabcolsep}{0.5pt}
\begin{tabular}{|c|c|c|c|c|}
\hline
\textbf{Schedules}&\textbf{V2C Update}&\textbf{C2V Update}&\textbf{Pre-computation}&\textbf{Comparison} \\\hline
\cline{1-5} 
\hline

\textbf{ BP} & \textbf{\textit{E}}& \textbf{\textit{E}}& \textbf{\textit{0}} & \textbf{\textit{0}} \\
\cline{1-5}
\textbf{ VFAP} & \textbf{\textit{E}}& \textbf{\textit{E}}& \textbf{\textit{0}} & \textbf{\textit{0}} \\
\cline{1-5}
\textbf{LBP} & \textbf{\textit{E($\hat{d_v}-1$)}}& \textbf{\textit{E}}& \textbf{\textit{0}}& \textbf{\textit{0}} \\
\cline{1-5}
\textbf{URW} & \textbf{\textit{E($\hat{d_v}-1$)}}& \textbf{\textit{E}}& \textbf{\textit{E($\hat{d_v}-1$)($\hat{d_c}-1$)}}& \textbf{\textit{E(E-1)}} \\
\cline{1-5}
\textbf{Rel.RBP } & \textbf{\textit{E($\hat{d_v}-1$) }}& \textbf{\textit{E}}& \textbf{\textit{E($\hat{d_v}-1$)($\hat{d_c}-1$)}}& \textbf{\textit{E(E-1)}} \\
\cline{1-5}
\textbf{RBP} & \textbf{\textit{E($\hat{d_v}-1$)}}& \textbf{\textit{E}}& \textbf{\textit{E(\textbf{$\hat{d_v}-1$})($\hat{d_c}-1$)}} & \textbf{\textit{E(E-1)}}\\
\cline{1-5}
\textbf{RD RBP 0.9} & \textbf{\textit{E($\hat{d_v}-1$)}}& \textbf{\textit{E}}& \textbf{\textit{E($\hat{d_v}-1$)($\hat{d_c}-1$)}} & \textbf{\textit{E(E-1)}}\\
\cline{1-5}
\textbf{SVNF} & \textbf{\textit{E(\textbf{$\hat{d_v}-1$})}}& \textbf{\textit{E}}& \textbf{\textit{E($\hat{d_v}-1$)($\hat{d_c}-1$)}} & \textbf{\textit{E($\hat{d_v}(\hat{d_c}-1)-1$)}}\\
\cline{1-5}
\textbf{WR-LBP} & \textbf{\textit{E($\hat{d_v}-1$)}}& \textbf{\textit{E}}& \textbf{\textit{E($\hat{d_v}-1$)($\hat{d_c}-1$)}} & \textbf{\textit{E(E-1)}}\\
\hline
\end{tabular}
\label{tab1}
\end{center}
\end{table*}

 \section{Simulations}

This section presents a simulation study illustrating the performance comparison of various algorithms including BP Flooding \cite{b4}, RBP \cite{b3}, RD \ RBP \ 0.9\cite{b6,b7}, Rel RBP \cite{b9}, SVNF  \cite{b5}, LBP \cite{b11}, URW \cite{b10}, VFAP \cite{b16}, and our proposed WR-LBP. The simulations utilize a parity check matrix $\mathbf{H}$ with dimensions $k = 256$ and $n = 512$, and a codeword length \textbf{$n$}. A regular LDPC code is employed, where each variable node and each check node possess identical degrees of connectivity. The code rate is set to $R=1/2$, and the number of blocks used to obtain Fig. 3. and Fig.4. is fixed at $5000$ frames. Fig. 3 employs $3$ decoding iterations whereas Fig. 4. employs $5$ iterations. Additionally, a BPSK modulation method is applied, transforming data from binary form $(0, 1)$ to waveforms $(-1, +1)$, which are then transmitted over an AWGN channel. Fig. 3. and Fig. 4. depict the comparison of Bit Error Rate (BER) versus the Signal-to-Noise Ratio ($E_b/N_0 \ dB$).

\begin{figure}[htbp]
\centerline{\includegraphics[width= 8.5 cm]{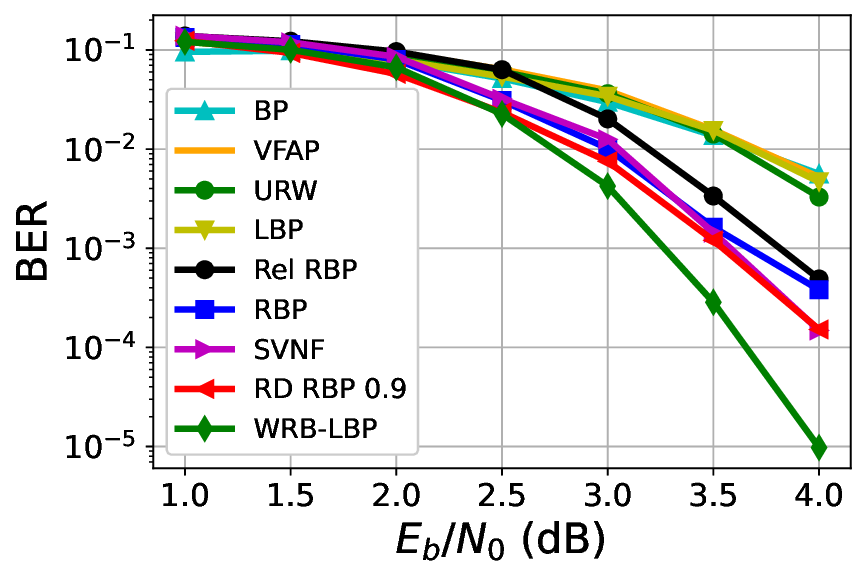}}
\caption{BER using 3 iterations and 5000 blocks. }
\label{fig}
\end{figure}

\begin{figure}[h]
\centering
\includegraphics[width=8.5 cm]{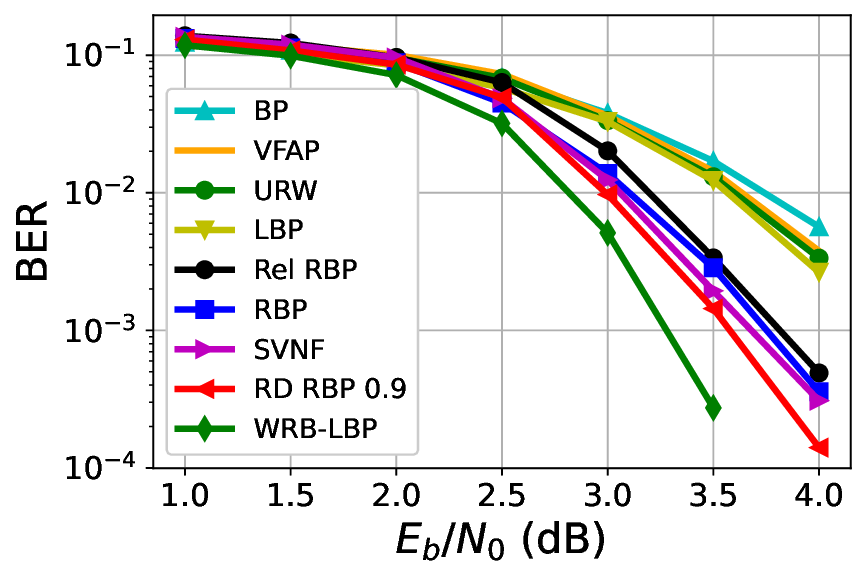}
\caption{BER using 5 iterations and 5000 blocks.}
\renewcommand{\figurename}{Figure}
\renewcommand{\footnotesize}{\fontsize{8}{10}\selectfont}
\end{figure} 

\begin{figure}[htbp]
\centerline{\includegraphics[width=8.5 cm]{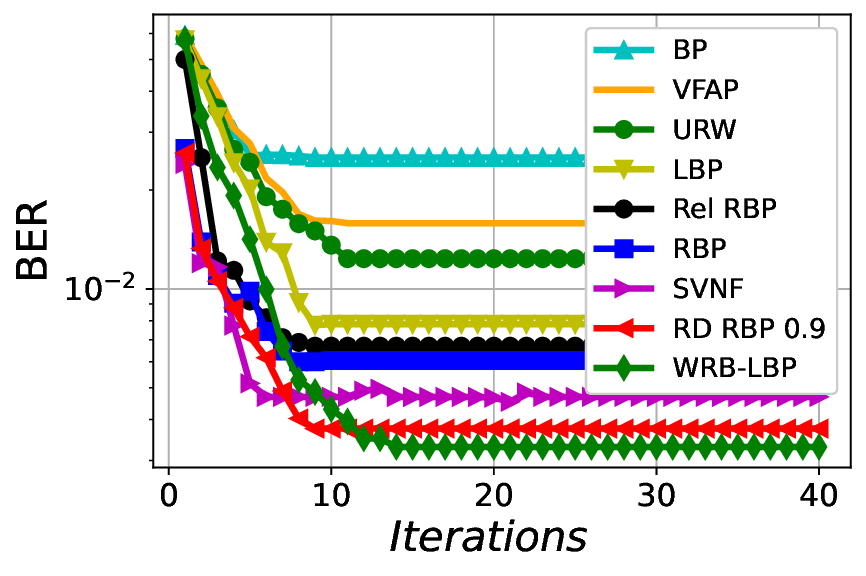}}
\caption{ BER versus the number of iterations with $E_b/N_0$=3.5 }
\label{fig}
\end{figure}

All algorithms commence with an initial SNR of 1 dB and remain consistent until reaching 1.5 dB. Beyond this threshold, the WR-LBP algorithm begins to exhibit superior performance in terms of BER compared to other algorithms. For instance, in Fig.3., at an SNR of $E_b/N_0 = 4$ dB, the algorithm closest to our proposal, RD \ RBP \ 0.9 , yields a BER of $1.48  \  10^{-4} dB $, whereas WR-LBP produces a BER of $9.765  \   10^{-4} dB$, resulting in a gain of approximately $1.38  \  10^{-4}$. Similarly, in Fig. 4. at an SNR of $E_b/N_0 = 3.5$ dB, WR-LBP outperforms other algorithms in decoding, achieving a gain of 0.001164 BER over the nearest competitor,  RD \ RBP \ 0.9.

In Fig. 5 we illustrate the BER corresponding to a specific Signal-to-Noise Ratio (SNR) of $E_b/N_0$ = 3.5 dB while varying the number of iterations from 1 to 40. The simulation reveals that our proposed method outperforms RD\ RBP  0.9, SVNF, RBP, and other approaches. However, it is notable that the convergence of our method is relatively slower within the initial 15 iterations. Nevertheless, beyond this point, our proposed technique demonstrates stable performance with a consistent BER.

 \section{Conclusion }
In this paper, we have introduced a novel IDS-type decoding algorithm, denoted as WR-LBP, by integrating a limited iteration count (5 and 3) and making use of a designated parity check matrix. The proposed WR-LBP algorithm operates within a structured layer framework for message passing. This methodology entails the arrangement of nodes into distinct layers, facilitating the systematic exchange of messages among adjacent layers. By orchestrating this organized communication flow, the proposed WR-LBP algorithm optimizes information dissemination and processing efficiency. Through extensive experimentation and comparative analysis, the WR-LBP algorithm demonstrates superior performance and complexity metrics when juxtaposed with competing algorithms in the field, establishing its efficacy and relevance in modern information processing systems.




\vspace{12pt}

\end{document}